\magnification=\magstep1
\overfullrule=0pt
\parskip=6pt
\baselineskip=15pt
\headline={\ifnum\pageno>1 \hss \number\pageno\ \hss \else\hfill \fi}
\pageno=1
\nopagenumbers
\hbadness=1000000
\vbadness=1000000

\input epsf

\vskip 25mm
\vskip 25mm
\vskip 25mm

\centerline{\bf ON POINCARE POLYNOMIALS OF HYPERBOLIC LIE ALGEBRAS }
\vskip 10mm

\centerline{\bf Meltem Gungormez} \centerline{Dept. Physics, Fac.
Science, Istanbul Tech. Univ.} \centerline{34469, Maslak, Istanbul,
Turkey } \centerline{e-mail: gungorm@itu.edu.tr}

\vskip 5mm

\centerline{\bf Hasan R. Karadayi} \centerline{Dept. Physics, Fac.
Science, Istanbul Tech. Univ.} \centerline{34469, Maslak, Istanbul,
Turkey } \centerline{e-mail: karadayi@itu.edu.tr} \centerline{and}
\centerline{Dept. Physics, Fac. Science, Istanbul Kultur University}
\centerline{34156, Bakirkoy, Istanbul, Turkey }

\medskip

\centerline{\bf{Abstract}}

We have general frameworks to obtain Poincare polynomials for Finite
and also Affine types of Kac-Moody Lie algebras. Very little is
known however beyond Affine ones, though we have a constructive
theorem which can be applied both for finite and infinite cases. One
can conclusively said that theorem gives the Poincare polynomial
P(G) of a Kac-Moody Lie algebra G in the product form P(G)=P(g) R
where g is a precisely chosen sub-algebra of G and R is a rational
function. Not in the way which theorem says but, at least for 48
hyperbolic Lie algebras considered in this work, we have shown that
there is another way of choosing a sub-algebra in such a way that R
appears to be the inverse of a finite polynomial. It is clear that a
rational function or its inverse can not be expressed in the form of
a finite polynomial.

Our method is based on numerical calculations and results are given
for each and every one of 48 Hyperbolic Lie algebras.

In an illustrative example however, we will give how above-mentioned
theorem gives us rational functions in which case we find a finite
polynomial for which theorem fails to obtain.

\hfill\eject

\vskip 3mm
\noindent {\bf{I.\ INTRODUCTION }}
\vskip 3mm

A characteristic fact about Poincare polynomial $P(G_N)$ of a
Kac-Moody Lie algebra {\bf [1]} $G_N$ of rank N is that the term of
an order s gives the number of its Weyl group elements which are
composed out of the products of s number of simple Weyl reflections
corresponding to simple roots of $G_N$ {\bf [2]}. For finite Lie
algebras, Poincare polynomials are known {\bf [3]} in the following
form

$$  P(G_N) = \prod^N_{i=1}  { t^{d_i}-1 \over t-1}  \eqno(I.1) $$

\noindent where $d_i$'s are the degrees of N basic invariants of
$G_N$ and t is always assumed to be an indeterminate. For an affine
Kac-Moody Lie algebra $\widehat{G}_N$ originated from a generic
finite Lie algebra $G_n$ where in general $N>=n$, Bott theorem {\bf
[4]} states that its Poincare polynomial has the following product
form

$$ P(\widehat{G}_N)=P(G_n) \prod^N_{i=1} {1 \over 1-t^{d_i-1}} \ .  \eqno(I.2) $$

Beyond Affine Kac-Moody Lie algebras, we only know a theorem
\footnote{$^{\dag}$}{p.123 of ref.3} which shows out a specific way
to obtain Poincare polynomial in any case, finite or infinite. It is
this theorem which in fact can be applied also in obtaining of (I.1)
and also (I.2). Theorem says that the Poincare polynomial $P(G_N)$
of a Kac-Moody Lie algebra $G_N$ has the product form
$$P(G_n)=P(g_n) \ R  \eqno(I.3) $$
\noindent where $P(g_n)$ is also the Poincare polynomial of a
sub-algebra $g_n\subset G_N $ with $N>n$. It is trivial to see that
theorem requires $g_n$ should be contained inside the Dynkin diagram
of $G_N$. Our concern in this work is the proposition of this
theorem which says that R is a {\bf rational function}. For a
Hyperbolic Lie algebra $ H_i $, our observation on the other hand is
that its Poincare polynomial comes in the form
$$  P(H_i)={P(G) \over Q_i(G) }   \eqno(I.4) $$
\noindent where $G$ is a properly chosen finite Lie Algebra and
$Q_i(G)$ is a polynomial of some finite degree in indeterminate t.
Due to the fact that a rational function or its inverse can only be
represented by a polynomial of infinite order, it is clear that
(I.3) and (I.4) say different and hence our reasoning in this work
is free of what the theorem says beyond Affine Kac-Moody Lie
algebras.

It is known that there are a finite number of Hyperbolic Lie
algebras {\bf [5]}. In the next section, we show how we obtain
Poincare polynomials for 48 one of them and hence the generalized
form (I.4). There are two points to mention here ;

{ \bf (1)} Degrees of polynomials $ Q_i(G) $ come in two different
values D or D-1 where D is the number of positive roots of the
chosen finite Lie algebra G.

{ \bf (2)} For the same $H_i$, there could be several equivalent
ways to specify the finite Lie algebra G so the explicit form of
denominator polynomials $ Q_i(G) $ depends on this choice. What is
important here is however that, on the contrary to above-mentioned
theorem,  there is no any requirement that G should be contained
inside the Dynkin diagram of $H_i$.

The following 3 examples will be instructive in all these points and
also reflect the basics of our calculational algorithm. In general
terms, let $W({G_N})$ be the Weyl group of $G_N$ and $ \sigma_i$'s
be its elements corresponding to simple roots $ \alpha_i$'s where
$i=1, \dots , N$ . In the following, we assume that reduced forms of
elements of a Weyl group can be expressed by the following notation:
$$ \Sigma(i_1, \dots ,i_k) \equiv \sigma_{i_1}. \dots \sigma_{i_k}. $$

Let $ S=\{\alpha_1, \dots , \alpha_6\} $ be the set of simple roots
of $H_{48}$ for which we use the following Dynkin diagram for the
first 2 examples:

\epsfxsize=6cm \centerline{\epsfbox{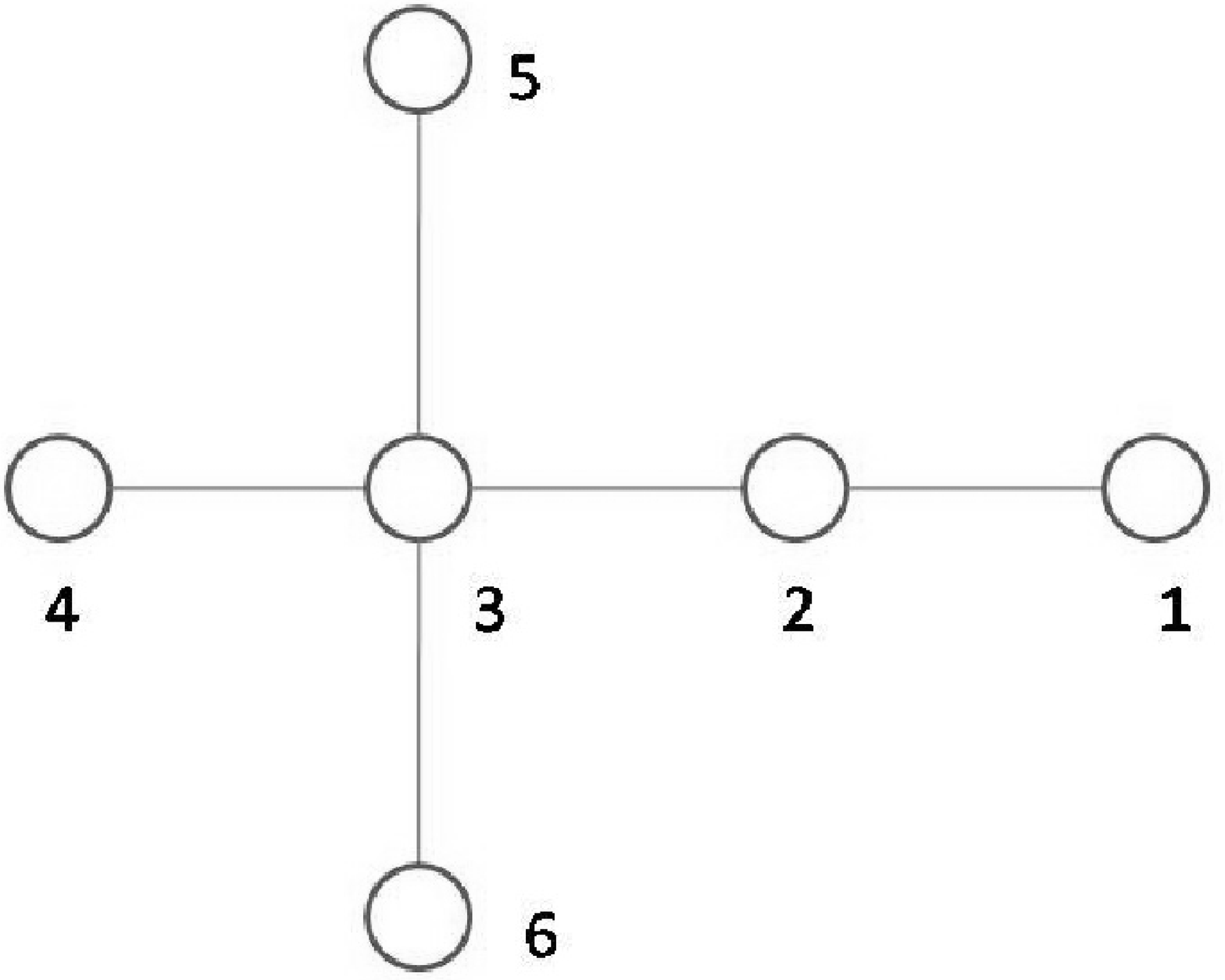}}

\noindent Among several possible choices, our 2 examples lead us
respectively to sub-algebras $ A_4 $ and $ D_5 $ of $H_{48}$, in
view of the following choices of 2 subsets of S:

\hfill\eject

\noindent $ (1) \ \  I_1 = \{\alpha_1, \dots , \alpha_4 \} \subset
S$

\noindent $ (2) \ \  I_2 = \{\alpha_1, \dots , \alpha_5 \} \subset
S$

\noindent At this point, one must notice that there could be no
choice for a subset which allows us to get a Lie sub-algebra which
is not contained inside the Dynkin diagram of $H_{48}$. It is now
seen for the first case that the above-mentioned theorem leads us to
a Poincare polynomial
$$ P(H_{48}) = P(A_4) \ R_1  \eqno(I.5) $$
where $P(A_4)$ is the Poincare polynomial of $A_4$ Lie algebra with
the Dynkin diagram

\epsfxsize=6cm \centerline{\epsfbox{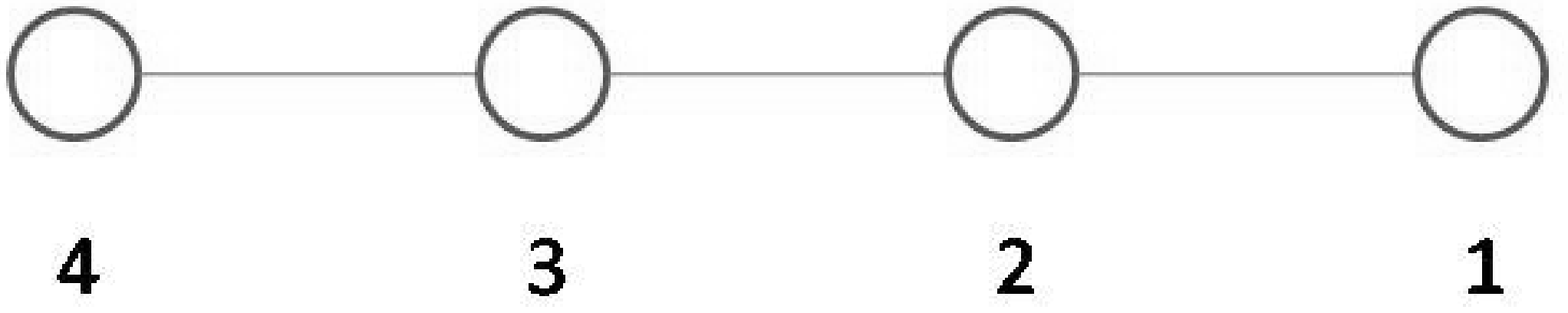}}

\noindent and $R_1$ is a rational function which we calculate here,
explicitly. In favor of (I.1), we know that
$$ P(A_4) =1 + 4 \ t + 9 \ t^2 + 15 \ t^3 + 20 \ t^4 + 22 \ t^5 + 20 \ t^6 + 15 \ t^7 +
 9 \ t^8 + 4 \ t^9 + t^{10}  $$
\noindent and corresponding 120 elements of $ W(H_{48})$ form a
subset $W({A_4}) \subset W(H_{48})$. Among infinite number of
elements of $ W(H_{48})$, there are only 120 elements originate only
from the set $ I_1 \subset S $ of example (1).

In (I.5), on the other hand, $R_1$ originates from the subset
$R_1(H_{48}) \subset W(H_{48}) $ for which any element $ w \in
W(H_{48})$ can be expressed by the factorized form
$$ w = u \ v \ \ , \ \ u \in W({A_4}) \ \ , \ \ v \in R_1(H_{48})  \eqno(I.6) $$
in such a way that
$$ \it{l}(w)= \it{l}(u) + \it{l}(v)   \eqno(I.7)  $$
where $ \it{l} $ is the length function of $H_{48}$. To exemplify
our algorithm, we give in the following the elements of
$R_1(H_{48})$ up to 6th order:

\hfill\eject

$$ \eqalign{
 &1 \cr
 &\sigma_5 , \sigma_6 , \cr
 &\Sigma(5,3) ,\Sigma(5,6) , \Sigma(6,3) , \cr
 &\Sigma(5,3,2) , \Sigma(5,3,4) , \Sigma(5,3,6) , \Sigma(5,6,3) , \Sigma(6,3,2) , \Sigma(6,3,4) , \Sigma(6,3,5) \cr
 &\Sigma(5,3,2,1) , \Sigma(5,3,2,4) , \Sigma(5,3,2,6) , \Sigma(5,3,4,6) , \Sigma(5,3,6,3) \Sigma(5,6,3,2) , \cr
 &\Sigma(5,6,3,4) , \Sigma(5,6,3,5) , \Sigma(6,3,2,1) , \Sigma(6,3,2,4) , \Sigma(6,3,2,5) , \Sigma(6,3,4,5) , \cr
 &\Sigma(5,3,2,1,4) , \Sigma(5,3,2,1,6) , \Sigma(5,3,2,4,3) , \Sigma(5,3,2,4,6) , \cr
 &\Sigma(5,3,2,6,3) , \Sigma(5,3,4,6,3) , \Sigma(5,3,6,3,2) , \Sigma(5,3,6,3,4) , \Sigma(5,3,6,3,5) , \cr
 &\Sigma(5,6,3,2,1) , \Sigma(5,6,3,2,4) , \Sigma(5,6,3,2,5) , \Sigma(5,6,3,4,5) , \Sigma(6,3,2,1,4) , \cr
 &\Sigma(6,3,2,1,5) , \Sigma(6,3,2,4,3) , \Sigma(6,3,2,4,5) , \Sigma(6,3,2,5,3) , \Sigma(6,3,4,5,3) , \cr
 &\Sigma(5,3,2,1,4,3) , \Sigma(5,3,2,1,4,6) , \Sigma(5,3,2,1,6,3) , \Sigma(5,3,2,4,3,5) , \cr
 &\Sigma(5,3,2,4,3,6) , \Sigma(5,3,2,4,6,3) , \Sigma(5,3,2,6,3,2) , \Sigma(5,3,2,6,3,4) , \cr
 &\Sigma(5,3,2,6,3,5) , \Sigma(5,3,4,6,3,2) , \Sigma(5,3,4,6,3,4) , \Sigma(5,3,4,6,3,5) , \cr
 &\Sigma(5,3,6,3,2,1) , \Sigma(5,3,6,3,2,4) , \Sigma(5,3,6,3,2,5) , \Sigma(5,3,6,3,4,5) , \cr
 &\Sigma(5,6,3,2,1,4) , \Sigma(5,6,3,2,1,5) , \Sigma(5,6,3,2,4,3) , \Sigma(5,6,3,2,4,5) , \cr
 &\Sigma(5,6,3,2,5,3) , \Sigma(5,6,3,4,5,3) , \Sigma(6,3,2,1,4,3) , \Sigma(6,3,2,1,4,5) , \cr
 &\Sigma(6,3,2,1,5,3) , \Sigma(6,3,2,4,3,5) , \Sigma(6,3,2,4,3,6) , \Sigma(6,3,2,4,5,3) , \cr
 &\Sigma(6,3,2,5,3,4) , \Sigma(6,3,2,5,3,6) , \Sigma(6,3,4,5,3,2) , \Sigma(6,3,4,5,3,6)    } $$
\noindent The reader could verify order by order that the number of
these elements do match with the first 6 terms in the infinite
polynomial expansion of the following rational function:
$$ R_1 ={(1 + t)^3 (1 + t^2) (1 - t + t^2) (1 + t^4) \over
1 - t^2 - 2 t^3 - t^4 + t^6 + t^7 + 3 t^8 + 2 t^9 - t^{13} - 2
t^{14} - 2 t^{15} - t^{16} + t^{19} + t^{20} } $$
\noindent Our
algorithm however allows us to investigate the existence of (I.5) at
any order. To this end, let us define
$$ P(H_{48}) \equiv \sum_{n=0}^\infty w_n \ t^n $$
and
$$ P(A_4) \equiv \sum_{n=0}^{10} u_n \ t^n   . $$
Since $R_1$ is a rational function, it could be represented also by
a polynomial of infinite order:
$$ R_1 \equiv \sum_{n=0}^\infty v_n \ t^n $$
The reader could also verify now that at any order $ M=0 , \dots ,
\infty $
$$ w_M = \sum_{s=0}^M u_s \ v_{M-s} . \eqno(I.8) $$

In case of example (2), one finds the following Dynkin diagram

\epsfxsize=6cm \centerline{\epsfbox{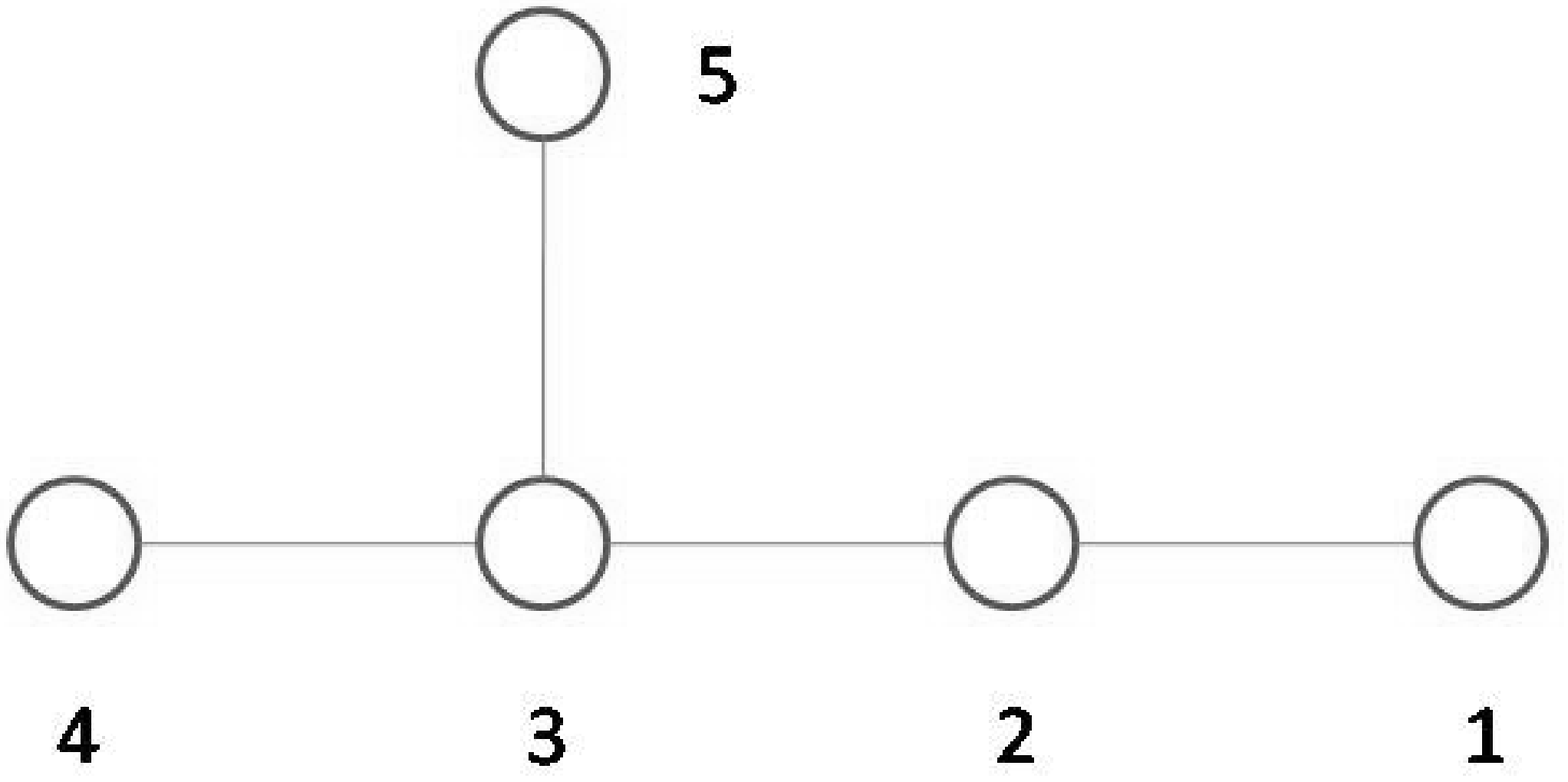}}

\noindent which gives us
$$ P(H_{48}) = P(D_5) \ R_2   \eqno(I.9) $$
where $ P(D_5) $ is the Poincare polynomial of $D_5$ Lie algebra. As
in the first example, $R_2$ is to be calculated in the form of the
following rational polynomial :
$$ R_2 ={ (1 + t) \over 1 - t^2 - 2 t^3 - t^4 + t^6 + t^7 + 3 t^8 + 2 t^9 - t^{13} - 2
t^{14} - 2 t^{15} - t^{16} + t^{19} + t^{20}}   $$

\noindent and the similar of (I.8) is seem to be valid.

For our last example, the Dynkin diagram of $H_{48}$ should be
defined, not in the way defined above but as in the following:

\epsfxsize=6cm \centerline{\epsfbox{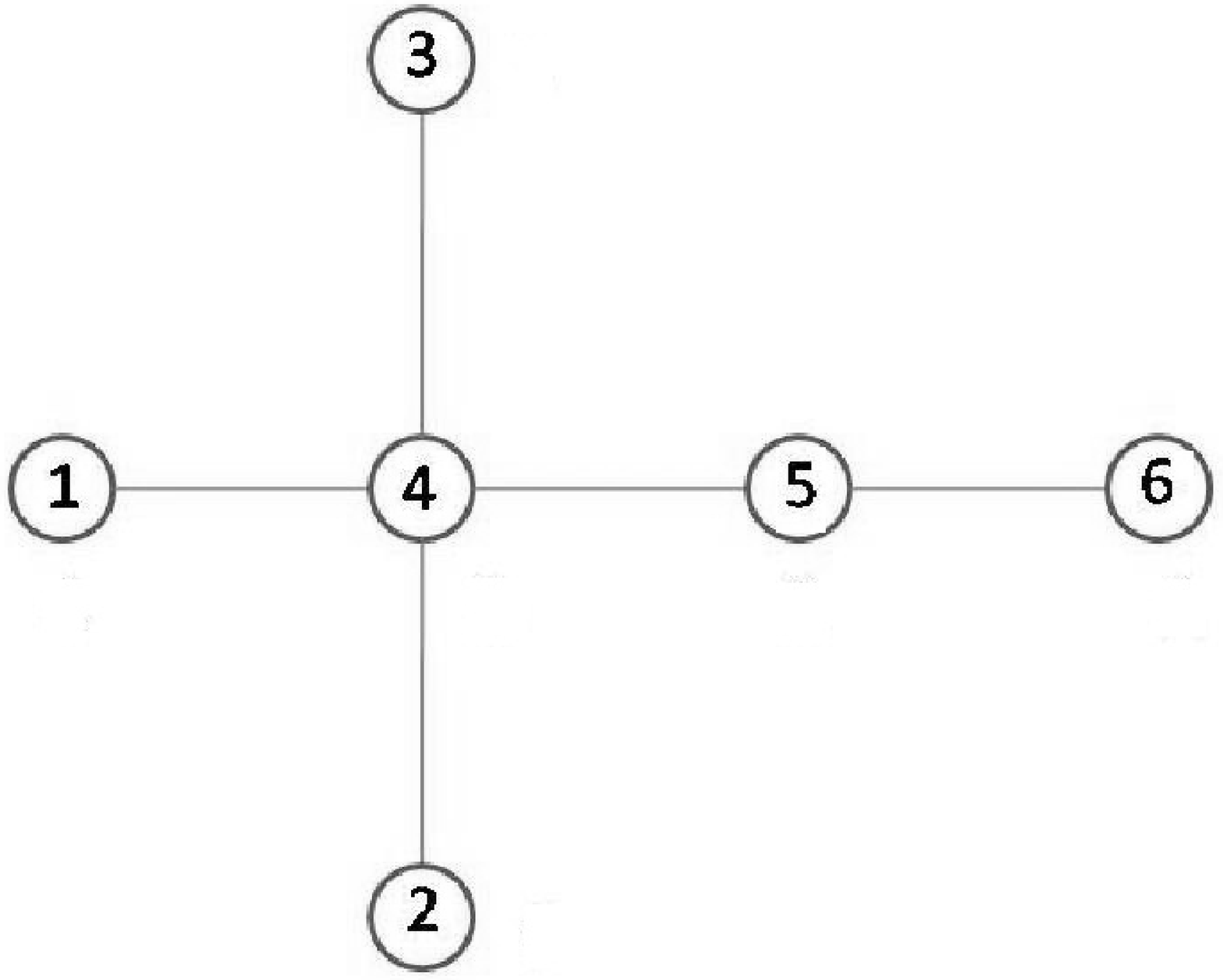}}

\noindent in such a way that the choice \noindent

\noindent $ (3) \ \  I_1 = \{\alpha_1, \dots , \alpha_5 \} \subset
S$

\noindent gives us an infinite sub-algebra which is in fact the
affine Lie algebra $ \widehat{D}_4 $ with the following Dynkin
diagram:

\epsfxsize=5cm \centerline{\epsfbox{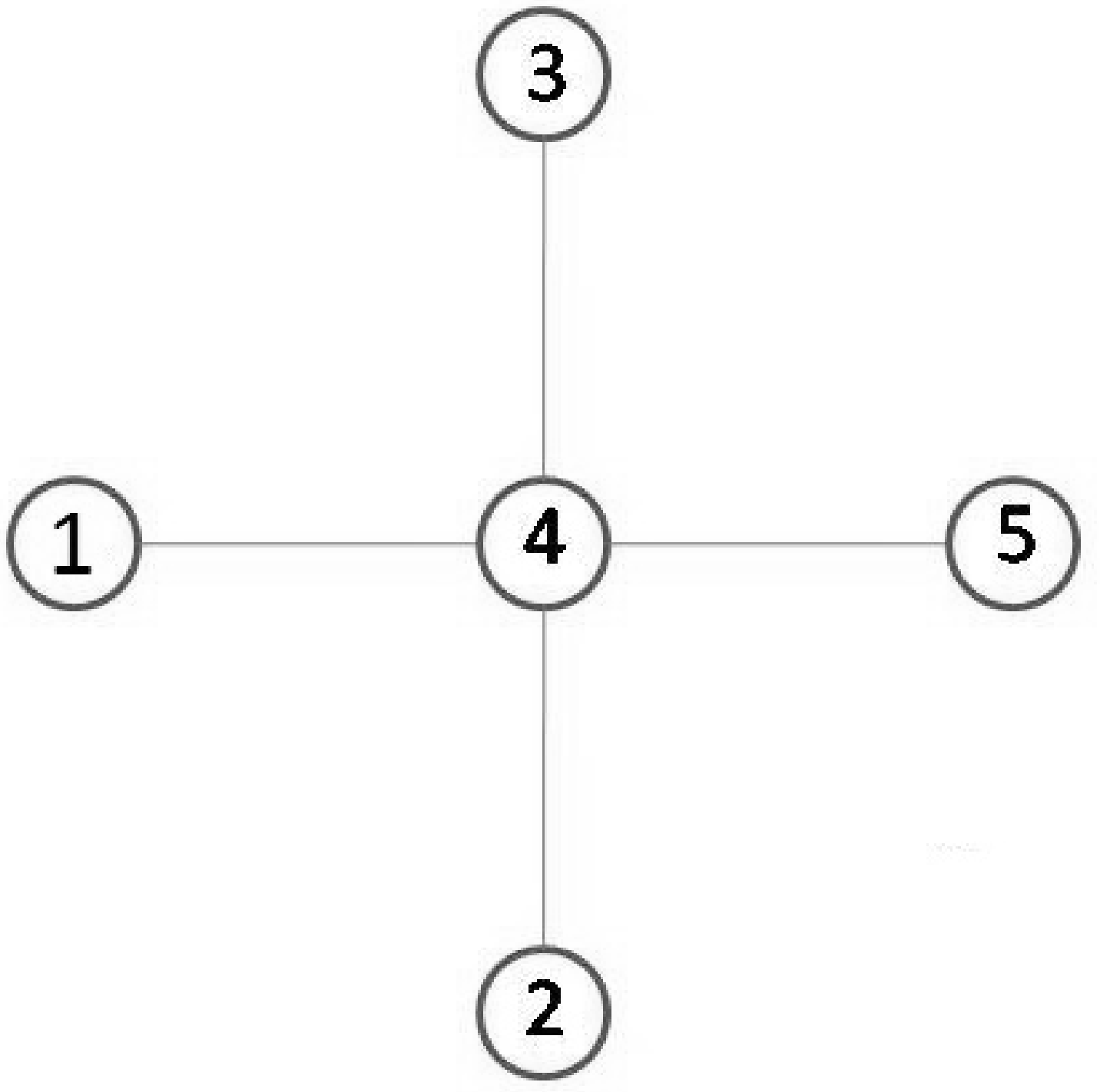}}

\noindent This time, similar of (I.5) and (I.9) is obtained in the
form

$$ P(H_{48}) = P(\widehat{D}_4) \ R_3   \eqno(I.10) $$
\noindent for which one has
$$ R_3 = {(1 - t)^3 (1 + t) (1 + t + t^2)^2 (1 + t^4) (1 + t + t^2 + t^3 + t^4) \over
 1 - t^2 - 2 t^3 - t^4 + t^5 + t^6 + t^8 + t^9 + t^{10} + t^{11} - t^{14} -
 t^{15} } . $$
\noindent In accordance with (I.2), with the notice that
$$ P(\widehat{D}_4) \equiv \sum_{n=0}^\infty u_n \ t^n , $$
\noindent one sees the similar of (I.8) is also valid here.

Let us conclude this section by giving the main motivation behind
this work against the well known existence of the above-mentioned
factorization theorem for Poincare polynomials, in general. As the
theorem said, in all examples corresponding polynomials $R_1, R_2$
and also $R_3$ are rational functions. In the next section, we show
however that, for Poincare polynomials of 48 Hyperbolic Lie
algebras, there is another factorization (I.4). We know from above
discussion that theorem could not give a way for obtaining such a
factorization.

\hfill\eject

\vskip 3mm \noindent {\bf{II.\ CALCULATION OF HYPERBOLIC POINCARE
POLYNOMIALS }} \vskip 3mm

We follow the notation of Kac-Moody-Wakimoto {\bf [6]} for
hyperbolic Lie algebras. To explain formal structure of our
calculations, we follow the example $ H_{48} $ from Appendix. Its
simple roots $\alpha_i$ and fundamental dominant weights $\lambda_i$
are given by equations
$$ \kappa(\lambda_i,\alpha_j) = \delta_{i,j} \ \ \ \ \ \  i,j = 1, \dots ,6. $$
where $\kappa( ,)$ is the symmetric scalar product which we know to
be exist, on $ H_{48}$ weight lattice, by its Cartan matrix A
$$ A = - \pmatrix{
 2&& -1&& 0&& 0&& 0&& 0&& \cr
 -1&& 2&& -1&& 0&& 0&& 0&& \cr
 0&& -1&& 2&& -1&& -1&& -1&& \cr
 0&& 0&& -1&& 2&& 0&& 0&& \cr
 0&& 0&& -1&& 0&& 2&& 0&& \cr
 0&& 0&& -1&& 0&& 0&& 2&& }$$
\noindent so we have
$$ \lambda_{i} = \sum_{j=1}^6 (A^{-1})_{i,j} \ \alpha_j $$

As in above, let $G_N$ be a chosen Kac-Moody Lie algebra, $ W(G_N)$
be its Weyl group and $\rho$ its Weyl vector. For any $\Sigma \in
W(G_N)$, let us now consider
$$ \Gamma \equiv \rho-\Sigma(\rho) \eqno(II.1)$$
which is by definition an element of the positive root lattice of
$G_N$. We know that $\Gamma$ is unique in the sense that  $\Gamma
\equiv \rho-\Sigma(\rho)$ is different from $\Gamma^\prime \equiv
\rho-\Sigma^\prime(\rho)$ for any two different $\Sigma$,
$\Sigma^\prime \in W(G_N)$. This could be easily understood due to
definition of Weyl vector which is in fact a strictly dominant
weight. This is sufficient to suggest our simple method to calculate
the number of Weyl group elements which are expressed in terms of
the same number of simple Weyl reflections $\sigma_i$ which are
defined by
$$ \sigma_i(\Lambda) \equiv \Lambda - 2 \
{ \kappa(\Lambda,\alpha_i) \over \kappa(\alpha_i,\alpha_i) } \
\alpha_i  \ \ , \ \ i =1,2,3, \dots $$ for any element $\Lambda$ of
weight lattice. Let us now consider k-tuple products
$$  \sigma_{i_1} \sigma_{i_2} \dots \sigma_{i_k}   \eqno(II.2) $$
which can not be reduced into products consisting less than k-number
of simple Weyl reflections, that is reduced elements. Out of all
these reduced elements as in (II.2), we define a class $ W^k \subset
W(G_N)$. The elements of any class $W^k$ are to be determined
uniquely by their actions on the Weyl vector which are different
from each other. We use a definitive algorithm to choose the ones
among the equivalents so, as is emphasized in the 3 examples given
above, our way of choosing the Weyl group elements originating from
the subsets (1), (2) and (3) is based on this algorithm. The aim of
this work doesn't need to show this algorithm here, in an explicit
way.

Now we can formally state that a Weyl group is the formal sum of its
classes $ W^k$. One should note that the number of elements of $ W^k
$ is always finite though the number of classes is finite for finite
and infinite for infinite Kac-Moody Lie algebras.

Looking back to $H_{48}$, we give some of its classes in the
following:
$$ \eqalign{
&W^0 = \{ 1 \} \cr &W^1 = \{ \sigma_1, \dots , \sigma_6 \} \cr &W^2
= \{ \Sigma(1,2) , \Sigma(1,3) , \Sigma(1,4) , \Sigma(1,5) ,
\Sigma(1,6) , \cr &\ \ \ \ \ \ \ \ \ \ \Sigma(2,1) ,  \Sigma(2,3) ,
\Sigma(2,4) , \Sigma(2,5) , \Sigma(2,6) , \cr &\ \ \ \ \ \ \ \ \ \
\Sigma(3,2) , \Sigma(3,4) , \Sigma(3,5) , \Sigma(3,6) , \Sigma(4,3)
, \cr &\ \ \ \ \ \ \ \ \ \ \Sigma(4,5) , \Sigma(4,6) , \Sigma(5,3) ,
\Sigma(5,6) , \Sigma(6,3) \} } $$
\noindent By setting $ \mid W^k
\mid $ to be the order of the set $W^k$, one has the polynomial $
\sum_{k=0}^\infty \ \mid W^k \mid \ t^k $ which is nothing but the
Poincare polynomial of $H_{48}$ . By explicit calculation up to 26th
order, we obtained the following result
$$ \eqalign{P(H_{48}) &\equiv
1 + 6 \ t^1 + 20 \ t^2 + 52 \ t^3 + 117 \ t^4 + 237 \ t^5 + 445 \
t^6 + 791 \ t^7 + 1347 \ t^8 + 2216 \ t^9 \cr &+ 3550 \ t^{10} +
5568 \ t^{11} + 8582 \ t^{12} + 13044 \ t^{13} + 19604 \ t^{14} +
29189 \ t^{15} \cr &+  43129 \ t^{16} + 63332 \ t^{17} + 92518 \
t^{18} + 134572 \ t^{19} + 195052 \ t^{20} + 281882 \ t^{21} \cr &+
406361 \ t^{22} + 584620 \ t^{23} + 839655 \ t^{24} + 1204232 \
t^{25} + \dots } \eqno(II.7)
$$ One sees that (II.7) is enough to conclude that
$$ P(H_{48}) \equiv { P(B_5) \over Q_{48}(B_5) } \eqno(II.8) $$
where
$$ Q_{48}(B_5) \equiv (1 - t - 2 \ t^3 + t^4 + t^6 - t^7 + 2 \
t^8 - t^9 + t^{10} + t^{12} + t^{13} - t^{14} - t^{15} -t^{18} -
t^{20} + t^{24}) \eqno(II.9) $$ and $P(B_5)$ is as given in (I.1)
for $B_5$ Lie algebra. As is emphasized in the first section, note
here that the number of positive roots of $B_5$ is equal to D=25 and
hence $Q_{48}(B_5)$ is a polynomial of order D-1=24.

The results of our calculations for 48 Hyperbolic Poincare
polynomials will be given in the following. Corresponding Dynkin
diagrams will also be given in Appendix.

$$ \eqalign{
&Q_1(B_2)=(1 - t - t^3 - t^4) \cr &Q_2(A_2)=(1-t-t^2-t^3) \cr
&Q_3(B_2)=(1 - t - t^2 - t^3 - 2 \ t^4) \cr &Q_4(B_2)=Q_1(B_2) \cr
&Q_5(A_2)=Q_2(A_2) \cr &Q_6(B_2)=Q_1(B_2)\cr &Q_7(B_2)=Q_3(B_2) \cr
&Q_8(A_3)=(1 - t - 2 t^2 + t^5 + t^6) \cr &Q_9(A_3)=(1 - t - 3 \ t^2
- t^4 + t^5 + 3 \ t^6)  \cr &Q_{10}(B_3)=(1 - t - 3 \ t^3 + t^4 -
t^5 + t^8 + 2 \ t^9) \cr &Q_{11}(B_3)=(1 - t - 2 \ t^3 + t^4 - t^5 +
t^8 + t^9) \cr &Q_{12}(B_3)=(1 - t - t^2 - 3 \ t^3 - t^6 + t^7 + t^8
+ 3 \ t^9) \cr &Q_{13}(B_3)=Q_{11}(B_3) \cr &Q_{14}(B_3)=Q_{11}(B_3)
\cr &Q_{15}(B_3)=Q_{10}(B_3) \cr &Q_{16}(G_2)=(1 - 2 \ t + t^2 - t^4
+ t^6) \cr &Q_{17}(B_3)=(1 - t - t^3 - 2 \ t^5 + t^6 + t^8 + t^9)
\cr &Q_{18}(G_2)=(1 - 2 \ t + t^3 - t^4 - t^5 + 2 \ t^6) \cr
&Q_{19}(B_3)=(1 - t - t^3 - t^4 - t^5 + t^6 + t^8 + t^9) \cr
&Q_{20}(G_2)=(1 - 2 \ t - 2 \ t^5 + 3 \ t^6) \cr &Q_{21}(D_4)=(1-t-2
\ t^2-t^3+2 \ t^5+t^6+2 \ t^7+3 \ t^8-t^9 +t^{10}-t^{11}-2 \ t^{12})
\cr &Q_{22}(B_4)=(1-t-t^2-t^3+t^7+2 \
t^8+t^9+t^{10}+t^{11}-t^{15}-t^{16}) \cr &Q_{23}(D_4)=(1 - t - t^3 -
t^4 + t^6 + t^7 + 2 \ t^8 - t^{11} - t^{12})  }
$$

$$ \eqalign{
&Q_{24}(G_2)=(1 - t - t^5) \cr &Q_{25}(B_2)=(1 - t - t^3) \cr
&Q_{26}(A_2)=(1-t-t^2) \cr &Q_{27}(B_2)=(1-t-t^2-t^3) \cr
&Q_{28}(G_2)=(1-t-t^3-t^5) \cr &Q_{29}(G_2)=(1-t-t^2-t^3-t^4-t^5)\cr
&Q_{30}(G_2)=Q_{24}(G_2) \cr &Q_{31}(B_2)=Q_{25}(B_2) \cr
&Q_{32}(B_2)=Q_{25}(B_2) \cr &Q_{33}(G_2)=Q_{28}(G_2) \cr
&Q_{34}(G_2)=Q_{28}(G_2) \cr &Q_{35}(B_2)=Q_{27}(B_2) \cr
&Q_{36}(A_3)=(1 - t - t^2 + t^5) \cr &Q_{37}(B_3)=(1 - t - t^3 +
t^8) \cr &Q_{38}(B_3)=(1 - t - t^2 - t^4 + t^6 + t^8) \cr
&Q_{39}(B_3)=(1 - t - t^2 - t^3 + t^7 + t^8) \cr &Q_{40}(B_3)=(1 - t
- 2 \ t^3 + t^4 - t^5 + t^6 + t^8) \cr &Q_{41}(B_3)=Q_{37}(B_3) \cr
&Q_{42}(B_3)=Q_{40}(B_3) \cr &Q_{43}(B_3)=(1 - t - t^4 + t^8) \cr
&Q_{44}(B_3)=(1 - t - t^3 - t^5 + t^6 + t^8) \cr
&Q_{45}(B_3)=Q_{44}(B_3) \cr &Q_{46}(D_5)=(1 - t^2 - t^3 - t^4 - t^5
- t^6 + t^8 + t^9 + 2 \ t^{10} + 2 \ t^{11} + 2 \ t^{12} + t^{13} +
t^{14} \cr &\ \ \ \ \ \ \ \ \ \ \ \ \ \ \ - t^{16} - t^{17} - t^{18}
- t^{19}) \cr &Q_{47}(D_5)=(1 - t - t^2 - t^4 + t^5 + t^6 + t^7 +
t^9 + t^{10} - t^{11} - t^{13} - t^{14} - t^{15} + t^{19}) \cr
&Q_{48}(B_5)=(1 - t - 2 t^3 + t^4 + t^6 - t^7 + 2 t^8 - t^9 + t^{10}
+ t^{12} + t^{13} - t^{14} \cr &\ \ \ \ \ \ \ \ \ \ \ \ \ \ \ \ -
t^{15} - t^{18} - t^{20} + t^{24} ) }
$$

\hfill\eject

\vskip 3mm
\noindent {\bf{III.\ CONCLUSION }}
\vskip 3mm

Let us conclude with the main idea of this work by the aid of a
beautiful example. One could say, for instance, that Bott theorem
and also the factorization theorem given above say the same thing.
Although this theorem proves useful as a calculational tool, Bott
theorem gives us a general framework to apply for affine Lie
algebras due to the fact that explicit calculation of a rational
function is in effect quite hard if it is not impossible. We note
again that a rational function can be expressed only in the form of
a polynomial of infinite order.

In the lack of such a general framework for Lie algebras beyond
affine ones, we also use an algorithm for explicit calculations.
Against the mentioned theorem, explicit calculations are possible
here due to the fact that in our formalism we only deal with
polynomials of some finite degree.

In a future publication, we also consider the case beyond Hyperbolic
Lie algebras, Kac-Moody Lie algebras of indefinite type.

\vskip3mm \noindent{\bf {REFERENCES}} \vskip3mm

\item [1] V. Kac, Infinite Dimensional Lie Algebras, Cambridge University Press, 1982
\item [2] J. E. Humphreys, Introduction to Lie Algebras and Representation Theory,
\item \ \ \ \ Springer-Verlag, 1972
\item [3] J. E. Humphreys, Reflection Groups and Coxeter Groups,
\item \ \ \ \ \ Cambridge University Press, 1990
\item [4] R. Bott, An Application of the Morse Theory to the topology of Lie-groups,
\item \ \ \ \ \ Bull. Soc. Math. France 84 (1956) 251-281
\item [5] C. Saclioglu, Dynkin Diagrams for Hyperbolic Kac-Moody Algebras,
\item \ \ \ \ J. Phys. A: Math. Gen. 22 (1989) 3753-3769  \ , \ See also [3]
\item [6] V. G. Kac, R. V. Moody and M. Wakimoto, On $E_{10}$, Differential Geometrical
\item \ \ \ \ Methods in Theoretical Physics, Kluwer Acad. Pub., 1988

\hfill\eject

\noindent {\bf{IV.\ APPENDIX }}
\vskip 10mm

\epsfxsize=12cm \centerline{\epsfbox{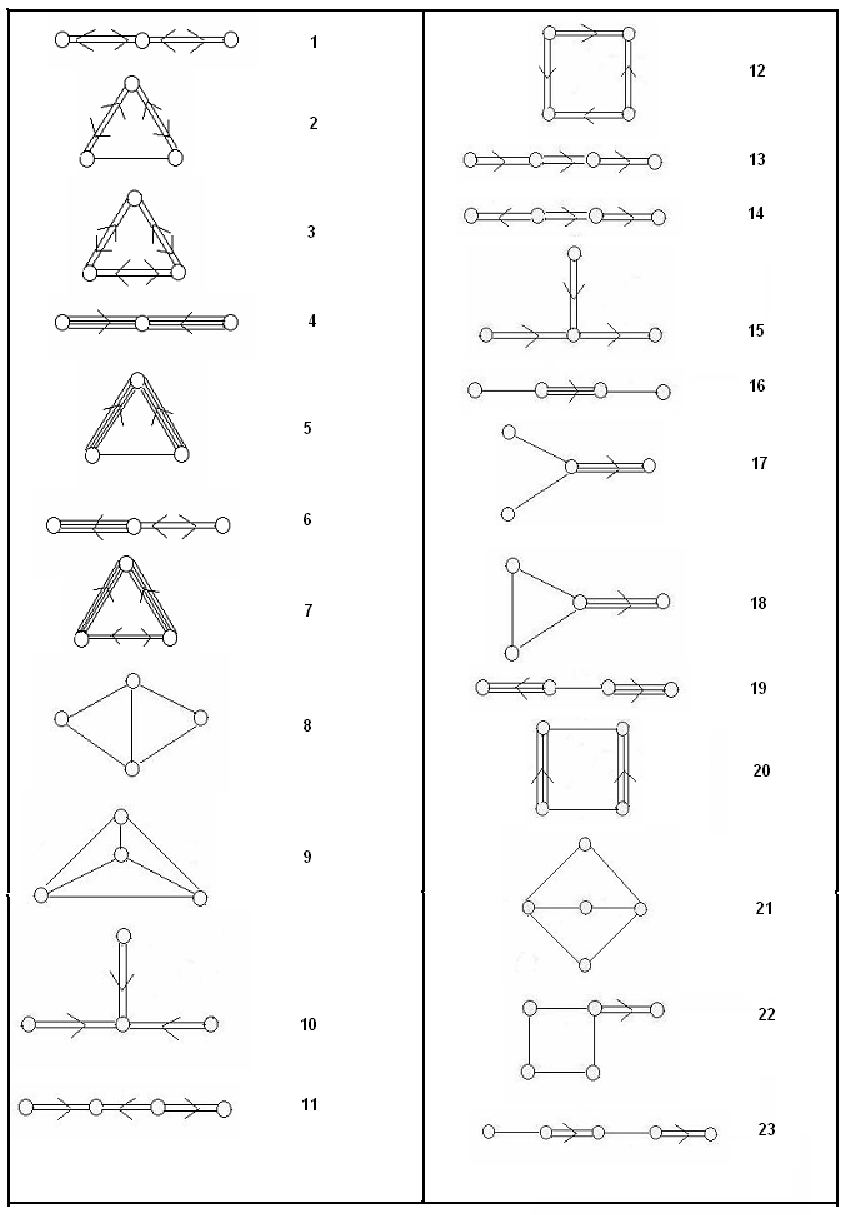}}

\epsfxsize=12cm \centerline{\epsfbox{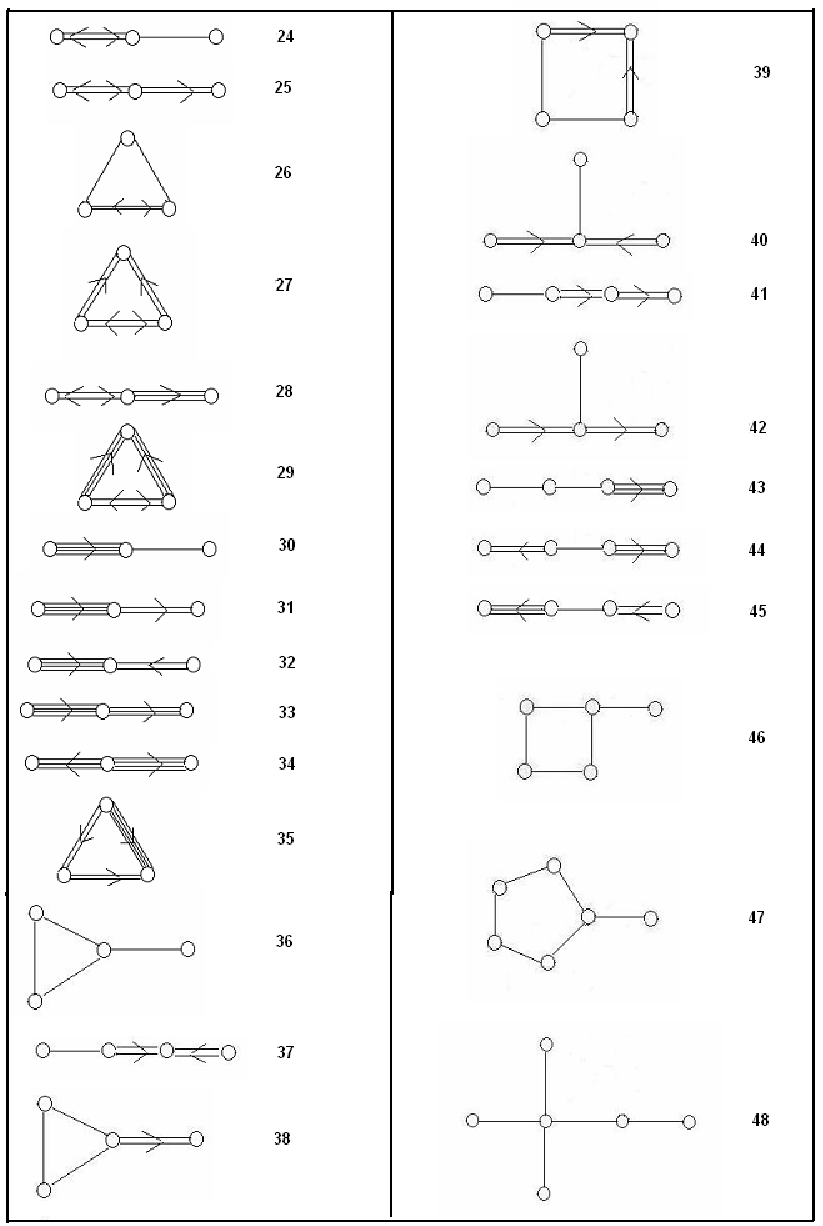}}

\end